# SLA-Based Coordinated Superscheduling Scheme and Performance for Computational Grids

Rajiv Ranjan, Aaron Harwood and Rajkumar Buyya
GRIDS Laboratory and P2P Group
Department of Computer Science and Software Engineering
University of Melbourne
Victoria, Australia
{rranjan, aharwood, raj}@cs.mu.oz.au

## Abstract

The Service Level Agreement (SLA) based grid superscheduling approach promotes coordinated resource sharing. Superscheduling is facilitated between administratively and topologically distributed grid sites by grid schedulers such as Resource brokers. In this work, we present a market-based SLA coordination mechanism. We based our SLA model on a well known *contract net protocol*.

The key advantages of our approach are that it allows: (i) resource owners to have finer degree of control over the resource allocation that was previously not possible through traditional mechanism; and (ii) superschedulers to bid for SLA contracts in the contract net with focus on completing the job within the user specified deadline. In this work, we use simulation to show the effectiveness of our proposed approach.

## 1 Introduction

A grid [23] enables aggregation of topologically distributed scientific instruments, storage structure, data, applications and computational resources. The notion of Grid computing is well beyond the traditional Parallel and Distributed Computing Systems (PDCS) as it involves various resources that belong to different administrative domains and are controlled by domain specific resource management policies. Furthermore, the grids have evolved around complex business and service models where various small sites (resource owners) collaborate for computational and economic benefits. The task of resource management and application scheduling over a grid is a complex undertaking due to resource heterogeneity, domain specific policy, dynamic environment, and various socio-economic and political factors.

The Grid superscheduling [37] problem is defined as: "scheduling jobs across the grid resources such as computational clusters, parallel supercomputers, desktop machines that belong to different administrative domains". Superscheduling in computational grids is facilitated by specialized superschedulers such as Grid Federation Agent [33], NASA-Superscheduler [38], Nimrod-G [6], and Condor-G [24]. Superscheduling activity involves (i) querying grid resource information services (GRIS) [14, 27, 19, 36, 8] for locating resources that match the job requirements; (ii) coordinating and negotiating SLAs; and (iii) job scheduling. The grid resources are managed by their local resource management systems (LRMSes) such as Condor [29], PBS [10], SGE [26], Legion [15], Alchemi [30] and LSF [4]. The LRMSes manage job queues, initiate and monitor their execution.

In this work, we propose a SLA [32, 20, 18] based coordinated superscheduling scheme for federated grid systems. An SLA is the agreement negotiated between a superscheduler (resource consumer) and LRMSes (resource provider) about acceptable job QoS constraints. These QoS constraints may include the job response time and budget spent. Inherently, a SLA is the guarantee given by a resource provider to a remote site job superscheduler for completing the job within the specified deadline or within the agreed budget or satisfying both at the same time. A SLA-based coordinated job superscheduling approach has many advantages: (i) It inhibits superschedulers from submitting unbounded amount of work to LRMSes; (ii) once a SLA is reached, users' are certain that agreed QoS shall be delivered by the system; (iii) job queuing or processing delay is significantly reduced thus leading to enhanced QoS, else a penalty model [48] is applied to compensate them ; and (iv) gives LRMSes more autonomy and better control over resource allocation decisions.

Our SLA model incorporates an economic mecha-nism [22, 12, 3, 5, 2] for job superscheduling and resource allocation. The economic mechanism enables the regulation of supply and demand of resources, offers incentive to the resource owners for leasing, and promotes QoS based resource allocation. We mainly focus on the decentralized commodity market model [47]. In this model every resource has a price, which is based on the demand, supply and value. An economy driven resource allocation methodology focuses on: (i) optimizing resource provider's payoff function; and (ii) increasing end-user's perceived QoS value. Note that our proposed superscheduling approach is studied as part of a new and emerging grid system which we call as Grid-Federation [33]. General details about this system can be found in section 2.

Our SLA model considers a collection of computational cluster resources as a contract net [41]. As job arrives, the grid superschedulers undertake one-to-one contract negotiation with the LRMSes managing the concerned resource. The SLA contract negotiation message includes: (i) whether a job can be completed within the specified deadline; and (ii) SLA bid expiration time (maximum amount of time a superscheduler is willing to wait before finalizing SLA). The SLA bid expiration time methodology we apply here is different from that adopted in the Tycoon system [28]. In Tycoon, the SLA bid expiration time at a resource is the same for all the jobs irrespective of their size or deadline. In this case, the total bid-processing delay is directly controlled by the local resource auctioneer. In our model, the superscheduler bids with a SLA bid expiration time proportional to the job's deadline. The focus is on meeting the job's SLA requirements, in particular the job deadline. The SLA contract negotiation in NASA-Superscheduler and Condor-Flock P2P [11] is based on general broadcast and limited broadcast communication mechanism respectively. Hence, these approaches have the following limitations: (i) high network overhead; and (ii) scalability problems.

Our time constrained SLA bid-based contract negotiation approach gives LRMSes finer control over the resource allocation decision as compared to traditional First-Come-First-Serve (FCFS approach). Existing superscheduling systems including NASA-Superscheduler, Condor-Flock P2P, Nimrod-G, Condor-G and Legion-Federation [46] assumes every LRMS allocates the resources using FCFS scheduling scheme. In this work, we propose a Greedy backfilling LRMS scheduling that focus on maximizing resource owner's payoff function. In this case, a LRMS maintains a queue of SLA bid requests generated by various superschedulers in the system at a time $t$. Every SLA bid has an associated expiry time with it. If the concerned LRMS does not reply within that period, then the SLA request is considered as expired. Greedy backfilling is based on well known Greedy or Knapsack method [25, 17, 9]. The LRMSes periodically iterates through the local SLA bids and finalizes the contract with those that fit the resource owner's payoff function.

The main contribution of this work includes (ii) SLA bid based superscheduling approach; (ii) a Greedy backfilling cluster scheduling approach for LRMSes that focus on maximizing the resource owners' payoff function; and (iii) allowing resource owners to have finer degree of control over resource allocation decisions. In this work, we use simulation to evaluate the feasibility of our proposed approach.

**Table 1. Notations**

| Symbol | Meaning |
|---|---|
| $n$ | number of GFAs. |
| $c_i$ | resource access cost at GFA $i$. |
| $p_i$ | number of processors at GFA $i$. |
| $J_{i,j,k}$ | $i$-th job from the $j$-th user of $k$-th GFA. |
| $p_{i,j,k}$ | number of processor required by $J_{i,j,k}$. |
| $b_{i,j,k}$ | assigned budget to $J_{i,j,k}$. |
| $d_{i,j,k}$ | assigned deadline to $J_{i,j,k}$. |
| $d^e_{i,j,k}$ | effective deadline for $J_{i,j,k}$. |
| $D(J_{i,j,k}, R_k)$ | time function (expected response time for $J_{i,j,k}$ at resource $k$). |
| $B(J_{i,j,k}, R_k)$ | cost function (expected budget spent for $J_{i,j,k}$ at resource $k$). |
| $I_k$ | incentive earned by resource owner $k$ over simulation period. |
| $\tau(J_{i,j,k})$ | returns next SLA bid interval $\Delta tneg_{i,j,k,p}$ for $J_{i,j,k}$. |
| $tneg_{i,j,k}$ | total SLA bid interval/delay for $J_{i,j,k}$. |
| $Q_{m,t}$ | set of jobs that have been assigned but not accepted at GFA $m$ at time $t$. |
| $Q^a_{m,t}$ | set of jobs that have been accepted at GFA $m$ at time $t$. |
| $Q^s_{m,t}$ | set of jobs sorted in decreasing order of incentive it provides to the resource owner at GFA $m$ at time $t$. |
| $n_u$ | number of users over all clusters ($\sum_{k=1}^{n} n_k$, $n_k$ number of users at GFA $k$). |
| $n_j$ | total jobs in the federation ($\sum_{(k,u_j)=1}^{n} n_k$, $u_i$). |
| $ts_{i,j,k}$ | job submission delay (user to GFA). |
| $tr_{i,j,k}$ | finished job return delay (GFA to user). |
| $\Delta tneg_{i,j,k,p}$ | total delay for $p$-th SLA bid for $J_{i,j,k}$. |
| $\lambda_{SLA_i}$ | SLA arrival rate at GFA $i$. |
| $\mu_{SLA_i}$ | SLA satisfaction rate at GFA $i$. |
| $l_{i,j,k}$ | job length for $J_{i,j,k}$ (in terms of million instructions) |
| $\alpha_{i,j,k}$ | communication overhead for $J_{i,j,k}$ |

## 1.1 Organization of this paper

The paper is organized as follows. In section 2, we present a brief overview of Grid-Federation superscheduling framework. Section 3.1 presents details about our proposed bid-based SLA contract negotiation model. In section 3.2, we give details about our proposed Greedy backfilling LRMS scheduling approach. In section 4, we present various experiments and discuss our results. Section 5 presents some of the related work in superscheduling. We end this paper with concluding remarks and our future vision in Section 6. Note that, in Table. 1 we define various notations that we use in this paper.

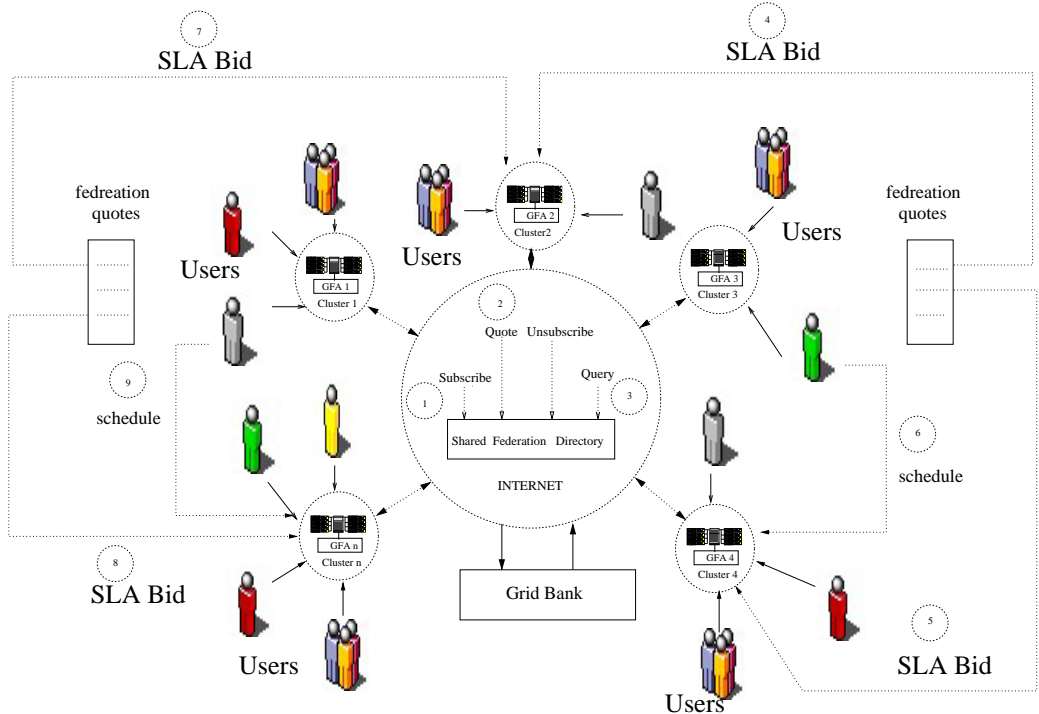

**Figure 1. Grid-Federation**

## 2 Brief overview of Grid-Federation

The Grid-Federation [33] system is defined as a large scale resource sharing system that consists of a cooperative federation of distributed clusters based on policies defined by their owners (shown in Fig.1). Fig.1 shows an abstract model of our Grid-Federation over a shared federation directory. To enable policy based transparent resource sharing between these clusters, we define and model a new RMS system, which we call Grid Federation Agent (GFA). Currently, we assume that the directory information is shared using some efficient protocol (e.g. a peer-to-peer protocol [31, 36, 8, 14]). In this case, the P2P system provides a decentralized database with efficient updates and range query capabilities. Individual GFAs access the directory information using the interface shown in Fig.1, i.e. subscribe, quote, unsubscribe, and query. The specifics of the interface can be found in [34]. Our approach considers the emerging computational economy metaphor [6, 43, 45] for the Grid-Federation. Some of the commonly used economic models [12] in resource allocation includes the commodity market model, the posted price model, the bargaining model, the tendering/contract-net model, the auction model, the bid-based proportional resource sharing model, the community/coalition model and the monopoly model. Grid-Federation considers decentralized commodity market model for managing job scheduling and resource allocation. In this case, the resource owners: (i) can clearly define what is shared in the Grid-Federation while maintaining a complete autonomy; (ii) can dictate who is given access; and (iii) get incentives for leasing their resources to federation users.

In Fig.1 a user who is local to GFA 3 is submitting a job. If the user's job QoS can't be satisfied locally then GFA 3 queries the federation directory to obtain the quote of the 1-st fastest (if the user is seeking optimize for time (OFT)) or 1-st cheapest cluster ( if the user is seeking optimize for cost (OFC)). In this case, the federation directory returns the quote advertised by GFA 2. Following this, GFA 3 bids for SLA contract (enquiry about QoS guarantee in terms of response time) at GFA 2. If GFA has too much load or the SLA bid does not fit the resource owner payoff function, the bid eventually timeouts. In this case, the SLA bid by GFA 2 timeouts. As next superscheduling iteration, GFA 3 queries the federation directory for the 2-nd cheapest/fastest GFA and so on. The process of SLA bids is repeated until GFA 3 finds a GFA that can schedule the job (i.e. accept the SLA bid) (in this example the job is finally scheduled on cluster 4).

# 3 Models

## 3.1 SLA model

The SLA model we consider is that of a set of distributed cluster resources each offering a fixed amount of processing power. The resources form part of the federated grid environment and are shared among the end-users each having its own SLA parameters. SLAs are managed and coordinated through admission control mechanism enforced by GFA at each resource site. Each user in the federation has a job $J_{i,j,k}$. We write $J_{i,j,k}$ to represent the $i$-th job from the $j$-th user of the $k$-th resource. A job consists of the number of processors required, $p_{i,j,k}$, the job length, $l_{i,j,k}$ (in terms of million instructions), the communication overhead, $\alpha_{i,j,k}$ and SLA parameters the budget, $b_{i,j,k}$, the deadline or maximum delay, $d_{i,j,k}$. More details about the job model can be found in [33].

### 3.1.1 SLA bid with expiration time (based on contract net protocol [41])

The collection of GFAs in the federation are referred to as a contract net and job-migration in the net is facilitated through the SLA contracts. Each GFA can take on two roles either a *manager* or *contractor*. The GFA to which a user submits a job for processing is referred as the manager GFA. The manager GFA is responsible for superscheduling the job in the net. The GFA which accepts the job from the manager GFA and overlooks its execution is referred to as the contractor GFA. Individual GFAs are not assigned these roles in advance. The role may change dynamically over time as per the user's job requirement. Thus, the GFA alternates between these two roles or adheres to both over the course of superscheduling.

As jobs arrive to a GFA, the GFA adopts the role of a manager. Following this, the manager GFA queries the shared federation directory to obtain the quote for the contractor GFA that matches the user specified SLA parameters. Note that, users can seek optimization for one of the SLA parameters i.e. either response time (OFT) or the budget spent (OFC). Once, the manager obtains the quote of the desired contractor, it undertakes one-to-one SLA contract negotiation with the contractor. The SLA contract negotiation message includes: (i) whether the job $J_{i,j,k}$ can be completed within the specified deadline; and (ii) SLA bid expiration time $\Delta t_{neg_{i,j,k,l}}$. The contractor GFA has to reply within the bid time $\Delta t_{neg_{i,j,k,l}}$, else the manager GFA undertakes SLA contract negotiation with the next available contractor in the net. Algorithm SLA bidding mechanism (refer to Algorithm 1) depicts various events and corresponding superscheduling actions undertaken by a GFA.

Our SLA contract model considers a part of the total job deadline as the SLA contract negotiation time (refer to Eq. 1). The manager GFA bids with different SLA expiration interval given by Eq. 2. In Fig. 2 we show the job superscheduling timeline. The timeline includes the job submission delay, $t_{s_{i,j,k}}$, total SLA contract negotiation delay, $t_{neg_{i,j,k}}$, expected response time (computed using Eq. 1) and finished job return delay, $t_{r_{i,j,k}}$. The total SLA contract bidding delay available to the manager GFA for superscheduling job $J_{i,j,k}$ is given by

$$t_{neg_{i,j,k}} = d_{i,j,k} - t_{s_{i,j,k}} - d^e_{i,j,k} - t_{r_{i,j,k}} \quad (1)$$

The total SLA contract bid negotiation delay $t_{neg_{i,j,k}}$ assumes a finite number of values $\Delta t_{neg_{i,j,k,1}}$, $\Delta t_{neg_{i,j,k,2}}$,...,$\Delta t_{neg_{i,j,k,n}}$ in superscheduling a job $J_{i,j,k}$ (refer to Fig. 2). We define the value of $\Delta t_{neg_{i,j,k,l}}$ by

$$\Delta t_{neg_{i,j,k,l}} = \frac{t_{neg_{i,j,k}} - \sum_{p=1}^{l-1} \Delta t_{neg_{i,j,k,p}}}{2} \; ; \quad l > 0 \quad (2)$$

Note that, the value for $\Delta t_{neg_{i,j,k,l}}$ can be given by other distributions [7] such as uniform or random. We intend to analyze various distributions for SLA bid interval and study its effect on our proposed superscheduling approach in our future work. For simplicity, in this work we use the distribution given by Eq. 2.

As the superscheduling iteration increases, the manager GFAs give less time to the contractor to decide on the SLA in order to meet the user's job deadline. This approach gives large number of scheduling iteration to the manager GFA. However, if the user's SLA parameters cannot be satisfied (after iterating up to the greatest $r$ such that GFA could feasibly complete the job) then the job is dropped. To summarize, a SLA bid for job $J_{i,j,k}$ includes

- $l$-th SLA bid expiry interval $t_{neg_{i,j,k,l}}$ (computed using Eq. 2);

- expected response time ($d^e_{i,j,k}$) (computed using Eq. 1).

We consider the function

$$\tau : J_{i,j,k} \longrightarrow \mathcal{Z}^+ \quad (3)$$

which returns the next allowed SLA bidding time interval $\Delta t_{neg_{i,j,k,p}}$ for a job $J_{i,j,k}$ using Eq.2.

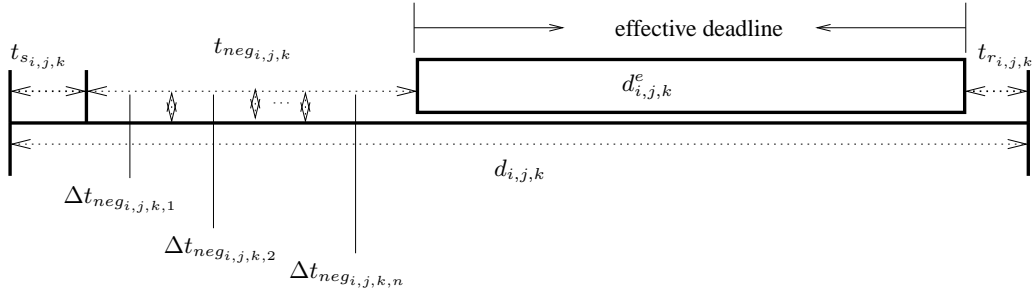

**Figure 2. Job superscheduling timeline**

**Algorithm 1**: SLA bidding mechanism

**0.1** PROCEDURE: SLA_BIDDING_MECHANISM
**0.2** **begin**
**0.3**    **begin**
**0.4**       SUB-PROCEDURE: EVENT_USER_JOB_SUBMIT ($J_{i,j,k}$)
**0.5**       call SLA_BID ($J_{i,j,k}$).
**0.6**    **end**
**0.7**    **begin**
**0.8**       SUB-PROCEDURE: SLA_BID ($J_{i,j,k}$)
**0.9**       Send SLA bid for job $J_{i,j,k}$ to the next available contractor GFA (obtained by querying the shared federation directory).
**0.10**   **end**
**0.11**   **begin**
**0.12**       SUB-PROCEDURE: EVENT_SLA_BID_REPLY ($J_{i,j,k}$)
**0.13**       **if** *SLA Contract Accepted* **then**
**0.14**          Send the job $J_{i,j,k}$ to accepting GFA.
**0.15**       **end**
**0.16**       **else**
**0.17**          call SLA_BID_TIMEOUT ($J_{i,j,k}$).
**0.18**       **end**
**0.19**   **end**
**0.20**   **begin**
**0.21**       SUB-PROCEDURE: SLA_BID_TIMEOUT($J_{i,j,k}$)
**0.22**       **if** $\tau(J_{i,j,k}) \geq 0$ **then**
**0.23**          call SLA_BID ($J_{i,j,k}$).
**0.24**       **end**
**0.25**       **else**
**0.26**          Drop the job $J_{i,j,k}$.
**0.27**       **end**
**0.28**   **end**
**0.29** **end**

### 3.2 Greedy backfilling: (LRMS scheduling model)

Most of the existing LRMSes apply system-centric policies for allocating jobs to the resources. Some of the well known system-centric policies include: (i) FCFS; (ii) Conservative backfilling [42]; and (ii) Easy backfilling [21]. Experiments [35] have shown that job backfilling approach offers significant improvement in performance over FCFS scheme. However, these system centric approaches allocate resource based on parameters that enhance system utilization or throughput. The LRMS either focuses on minimizing the response time (sum of queue time and actual execution time) or maximizing overall resource utilization of the system and these are not specifically applied on a per-user basis (user oblivious). Further, the system centric LRMSes treat all resources with the same scale, thus neglecting the resource owner payoff function. In this case, the resource owners do not have any control over resource allocation decisions. While in reality the resource owner would like to dictate how his resources are made available to the outside world and apply a resource allocation policy that suits his payoff function. To summarize, the system-centric approaches do not provide mechanisms for resource owners to dictate resource: (i) sharing; (ii) access and ; (iii) allocation policies.

To address this, we propose a Greedy method based resource allocation heuristic for LRMSes. Our proposed heuristic focuses on maximizing payoff function for the resource owners. The heuristic is based on the well known Greedy method [9, 17]. The Greedy method for solving optimization problems considers greedily maximizing or minimizing the short-term goals and hoping for the best without regard to the long-term effects. This method has been used to solve *the knapsack problem* [25]. Given a set $S$, of $n$ items, with each item $i$, having a positive benefit $b_i$, a positive weight $w_i$, knapsack capacity $W$ and amount $x_i$, we consider for each item $i$. The Greedy heuristic focuses on maximizing total benefit $\sum_{i \in S^a} b_i(x_i/w_i)$ with constraint $\sum_{i \in S^a} x_i \leq W$, such that $S^a \subseteq S$.

Fig.3 shows the queue of SLA bids at each site in the federation. Every incoming SLA bid is added to the LRMS request queue, $Q_{m,t}$ and a bid expiration timeout event is scheduled after time interval $\tau(J_{i,j,k})$. Every resource $i$ has different SLA bid arrival rate, $\lambda_{SLA_i}$ and SLA bid satisfaction rate, $\mu_{SLA_i}$. The LRMS scheduler iterates through the SLA bid queue in case any of the following events occur: (i) new SLA bid arrives to the site; (ii) job completion; or (iii) SLA bid reaches its expiration time. Procedure Greedy backfilling (refer to

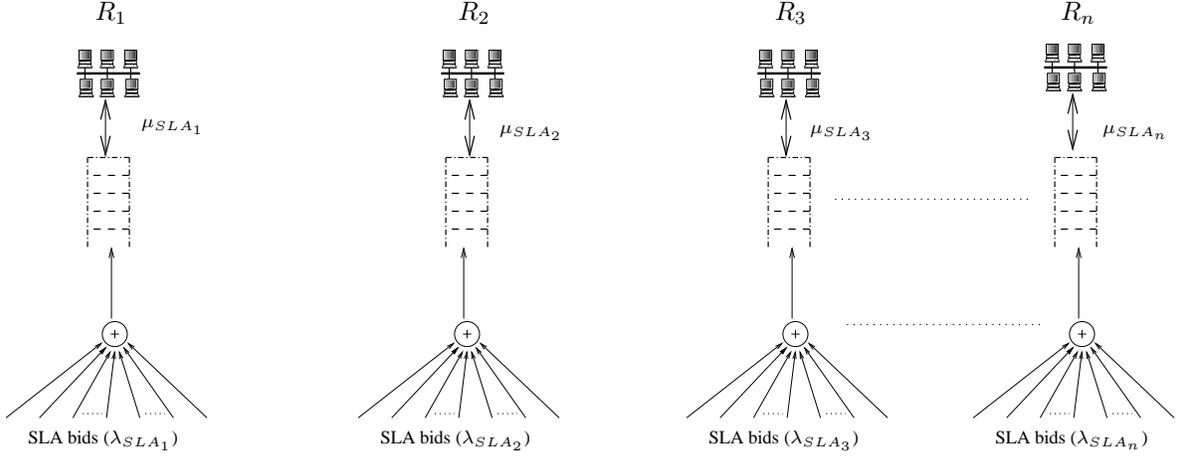

**Figure 3. SLA bid queues in Grid-Federation**

Algorithm 2) depicts various events and corresponding scheduling actions undertaken by the LRMS.

### Integer linear programming (ILP) formulation of scheduling heuristic

Queue, $Q_{m,t}$, maintains the the set of job SLA bids currently negotiated with the LRMS at GFA $m$ by time $t$. We consider the SLA bid acceptance variable $x_{i,j,k}$

### Definition of variable

$x_{i,j,k} = 1$ if the SLA request for job $J_{i,j,k}$ is accepted;
$x_{i,j,k} = 0$ otherwise.

The Greedy-Backfilling heuristic accepts SLA requests constrained to the availability of number of processors requested for job $J_{i,j,k}$ and expected response time $d^e_{i,j,k}$.

### Definition of the constraints

$$\sum_{\substack{1 \leq i \leq n_j \\ 1 \leq j \leq n_u \\ 1 \leq k \leq n}} p_{i,j,k} \leq p_m \quad (4)$$

$p_m$ total number of processors available at a LRMS (GFA) $m$. $p_{i,j,k}$ denotes number of processor requested by the SLA bid for job $J_{i,j,k}$. All the accepted SLA bids for jobs are maintained in the queue $Q^a_{m,t}$.

### Payoff or Objective function

The LRMS scheduler accepts SLA bids for the jobs such that it maximizes the resource owners' payoff function by applying Greedy backfilling heuristic

$$I_m = max(\sum_{\substack{1 \leq i \leq n_j \\ 1 \leq j \leq n_u \\ 1 \leq k \leq n \\ 1 \leq m \leq n}} B(J_{i,j,k}, R_m)) \quad (5)$$

## 3.3 Economic parameters

### 3.3.1 Setting price ($c_i$)

The resource owners configure the resource access cost $c_i$ to reflect its demand in the federation. A resource owner can vary $c_i$ depending on resource demand $\lambda_{SLA_i}$ and resource supply $\mu_{SLA_i}$ pattern. In case, $\lambda_{SLA_i} > \mu_{SLA_i}$, then the resource owner can increase the $c_i$. However, $\lambda_{SLA_i}$ depends on the user population profile. If the majority of users are seeking optimization for response time then time-efficient resources may increase $c_i$ until $\lambda_{SLA_i} = \mu_{SLA_i}$. Furthermore, to find the bounds of $c_i$ it is mandatory to consider the amount of budget available to the users.

For simplicity, in this work we assume that $c_i$ remains static throughout the simulations. We intend to analyze different pricing algorithm [16, 39, 47] based on supply and demand function as a future work. Using the static price $c_i$, we quantify how varying the SLA bid time affects the federated superscheduling systems' performance. In simulations, we configure $c_i$ using the function:

$$c_i = f(\mu_i) \quad (6)$$

where,

$$f(\mu_i) = \frac{c}{\mu} \mu_i \quad (7)$$

**Algorithm 2**: Greedy-Backfilling

```
1.1  PROCEDURE: GREEDY_BACKFILLING
1.2  begin
1.3      r = p_m
1.4      c = 0
1.5      Q_{m,t} ← φ
1.6      Q^a_{m,t} ← φ
1.7      Q^s_{m,t} ← φ
1.8      begin
1.9          SUB-PROCEDURE:Event_SLA_Bid_ARRIVAL(J_{i,j,k})
1.10         A SLA request message for the job J_{i,j,k} that arrives at a GFA Q_{m,t} ← Q_{m,t} ∪ {J_{i,j,k}}
1.11         Schedule the SLA bid timeout event after τ(J_{i,j,k}) time units
1.12         call STRICT_GREEDY()
1.13     end
1.14     begin
1.15         SUB-PROCEDURE:Event_SLA_Bid_Timeout(J_{i,j,k})
1.16         A SLA bid for job J_{i,j,k} that reaches timeout period
1.17         if (r ≥ p_{i,j,k} and d^e_{i,j,k} ≥ D(J_{i,j,k}, R_m)) then
1.18             Call RESERVE(J_{i,j,k})
1.19         end
1.20         else
1.21             Reject the SLA bid for job J_{i,j,k}
1.22             Reset Q_{m,t} ← Q_{m,t} − {J_{i,j,k}}
1.23         end
1.24     end
1.25     begin
1.26         SUB-PROCEDURE:Event_Job_Finish(J_{i,j,k})
1.27         A job J_{i,j,k} that finishes at a GFA Reset r = r + p_{i,j,k}
1.28         call STRICT_GREEDY()
1.29     end
1.30     begin
1.31         SUB-PROCEDURE: RESERVE(J_{i,j,k})
1.32         Reserve p_{i,j,k} processors for the job J_{i,j,k}
1.33         Reset r = r − p_{i,j,k}, Q_{m,t} ← Q_{m,t} − {J_{i,j,k}}, Q^a_{m,t} ← Q^a_{m,t} ∪ {J_{i,j,k}}
1.34     end
1.35     begin
1.36         SUB-PROCEDURE: STRICT_GREEDY()
1.37         Reset c = 0
1.38         Sort SLA bids in Q_{m,t} in decreasing order of incentives and store in Q^s_{m,t}
1.39         Get next SLA bid for job J_{i,j,k} from the list Q^s_{m,t}, c=c+1
1.40         if (r ≥ p_{i,j,k} and d^e_{i,j,k} ≥ D(J_{i,j,k}, R_m)) then
1.41             Call RESERVE(J_{i,j,k})
1.42         end
1.43         else
1.44             if c < sizeof(Q^s_{m,t}) then
1.45                 Iterate through step 1.39
1.46             end
1.47         end
1.48     end
1.49 end
```

$c$ is the access price and $\mu$ is the speed of the fastest resource in the Grid-Federation. Details about how users are charged on per job basis can be found in [33].

### 3.3.2 User budget and deadline

While our simulations in the next section use trace data for job characteristics, the trace data does not include user specified budgets and deadlines on a per job basis. In order to study our proposed SLA model and superscheduling approach, we are forced to fabricate these quantities and we include the models here.

For a user, $j$, we allow each job from that user to be given a budget,

$$b_{i,j,k} = 2\, B(J_{i,j,k}, R_k). \qquad (8)$$

In other words, the total budget of a user over simulation is unbounded and we are interested in computing the budget that is required to schedule all of the jobs.

Also, we let the deadline for job $i$ be

$$d_{i,j,k} = 3\, D(J_{i,j,k}, R_k). \qquad (9)$$

We assign three times the expected response time for the given job, as compared to expected response time on the originating resource. We use the multiplying constant as 3, for allowing the superschedulers ample time during SLA bidding. However, as a future work we intend to analyze how does the system performance changes when multiplying constant approaches 1 and infinity. Details about the budget and time function can be found in [33].

## 4 Experiments and observations

### 4.1 Workload and resource methodology

We performed trace based simulation to evaluate the effectiveness of our SLA based superscheduling approach. The workload trace data was obtained from [1]. The trace contains real time workload of various resources/supercomputers that are deployed at the Cornell Theory Center (CTC SP2), Swedish Royal Institute of Technology (KTH SP2), Los Alamos National Lab (LANL CM5), LANL Origin 2000 Cluster (Nirvana) (LANL Origin), NASA Ames (NASA iPSC) and San-Diego Supercomputer Center (SDSC Par96, SDSC Blue, SDSC SP2) (See Table 2). The workload trace is a record of usage data for parallel jobs that were submitted to various resource facilities. Every job arrives, is allocated one or more processors for a period of time, and then leaves the system. Furthermore, every job in the workload has an associated arrival time, indicating

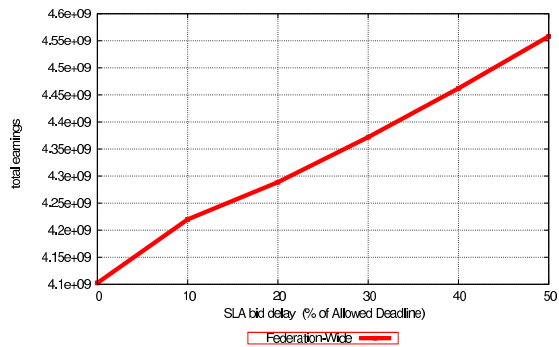

(a) total SLA bid delay vs. total federation earning (grid dollars)

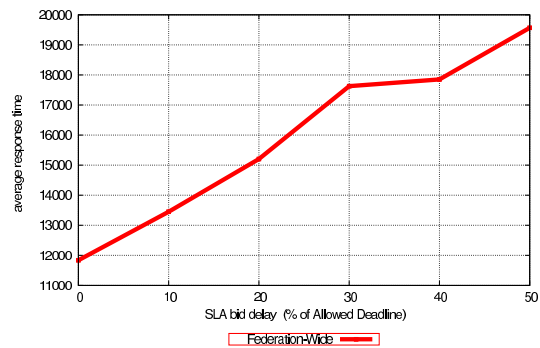

(b) total SLA bid delay vs. average response time (sim units)

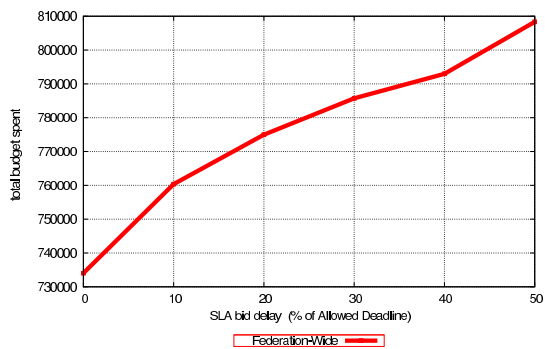

(c) total SLA bid delay vs. average budget spent (grid dollars)

**Figure 4. Federation perspective**

**Table 2. Workload and Resource Configuration**

| Index | Resource / Cluster Name | Trace Date | Processors | MIPS (rating) | Jobs | Quote(Price) | NIC to Network Bandwidth (Gb/Sec) |
|---|---|---|---|---|---|---|---|
| 1 | CTC SP2 | June96-May97 | 512 | 850 | 79,302 | 4.84 | 2 |
| 2 | KTH SP2 | Sep96-Aug97 | 100 | 900 | 28,490 | 5.12 | 1.6 |
| 3 | LANL CM5 | Oct94-Sep96 | 1024 | 700 | 201,387 | 3.98 | 1 |
| 4 | LANL Origin | Nov99-Apr2000 | 2048 | 630 | 121,989 | 3.59 | 1.6 |
| 5 | NASA iPSC | Oct93-Dec93 | 128 | 930 | 42,264 | 5.3 | 4 |
| 6 | SDSC Par96 | Dec95-Dec96 | 416 | 710 | 38,719 | 4.04 | 1 |
| 7 | SDSC Blue | Apr2000-Jan2003 | 1152 | 730 | 250,440 | 4.16 | 2 |
| 8 | SDSC SP2 | Apr98-Apr2000 | 128 | 920 | 73,496 | 5.24 | 4 |

when it was submitted to the scheduler for consideration. As the experimental trace data does not include details about the network communication overhead involved for different jobs, we artificially introduced the communication overhead element as 10% of the total parallel job execution time. More details about the job execution time modeling can be found in [33]. The simulator was implemented using GridSim [13] toolkit that allows modeling and simulation of distributed system entities for evaluation of scheduling algorithms. To enable the parallel workload simulation with GridSim, we extended the existing GridSim's Alloc Policy and Space Shared entities.

Our simulation environment models the following basic entities in addition to existing entities in GridSim:

- local user population – models the workload obtained from trace data;

- GFA – generalized RMS system;

- GFA queue – placeholder for incoming jobs from local user population and the federation;

- GFA shared federation directory – simulates an efficient distributed query process such as peer-to-peer.

For evaluating the SLA based superscheduling, we assigned a synthetic QoS specification to each resource including the Quote value (price that a cluster owner charges for service), with varying MIPS rating and underlying network communication bandwidth. The simulation experiments were conducted by utilizing workload trace data over the total period of four days (in simulation units) at all the resources. We consider the following resource sharing environment for our experiments:

- federation with economy – Experiments 1 and 2.

### 4.2 Experiment 1 - Quantifying scheduling parameters related to resource owners and end-users with varying total SLA bid time

We quantify the following scheduling parameters related to resource owners and end-users

- resource owner: payoff function (total earnings, earnings per processor), resource utilization (in terms of total MI executed);

- end-users: QoS satisfaction (average response time, average budget spent), number of jobs accepted.

We performed the simulations which comprised of end-users seeking OFT for their jobs (i.e. 100% users seek OFT). We vary the total SLA bid from 0% to 50% of total allowed job deadline. In case, no SLA bid delay is allowed (i.e. 0% of total allowed deadline) then the contacted GFA has to instantly make the admission control decision. In this case, we simulate FCFS based strategy for finalizing SLA. However, in other cases we consider Greedy backfilling SLA approach.

### 4.3 Experiment 2 - Quantifying message complexity involved with varying total SLA bid time

In this experiment we consider the message complexity involved in our proposed superscheduling approach. In this work, we consider the following superscheduling parameters related to overall system message complexity:

- average message per job at a resource $i$: the number of SLA bid requests undertaken at a resource on the average before the job was actually scheduled ;

- local message count: number of SLA bid scheduling messages undertaken for local jobs at a resource $i$;

- remote message count: number of SLA bid scheduling message overhead for remote jobs at a resource $i$.

## 4.4 Results and observations

### 4.4.1 Federation perspective

In experiment 1, we measure how the varying of the total time for SLA bids coupled with Greedy backfilling resource allocation strategy affects the Grid participants across the federation. We quantify how the additional decision making time given to the LRMSes before finalizing the SLA contracts affects the overall system performance in terms of resource owner's and end-user's objective functions. We observed that when the LRMSes across the federation applied FCFS technique for finalizing the SLAs (i.e. no decision making time was given, so the LRMSes have to reply as soon as the SLA request was made), the resource owner's made $4.102 \times 10^9$ grid dollars as incentive (refer to Fig.4(a)).

We observed that with an increase in the total SLA bidding time (i.e. as the LRMSes were allowed decision making time before finalizing the SLAs hence they applied Greedy backfilling scheduling on the queue of SLA bids), the resource owners earned more incentive as compared to FCFS case. When 10% of total deadline was allowed for SLA bids, the total incentive earned across the federation increased to $4.219 \times 10^9$ grid dollars. While, in case 50% of total job deadline was allowed for the SLA bids, the total incentive accounted to $4.558 \times 10^9$ grid dollars. Hence, the resource owners across federation exprienced an increase of approximately 10% in their incentives as compared to the FCFS case.

However, we observed that with an increase in the total SLA bid delay, the end-users across the federation experienced degraded QoS. During the FCFS case, the average response time across the federation was $1.183 \times 10^4$ sim units (refer to Fig.4(b)). However, in case of 10% SLA bid delay the average response time increased to $1.344 \times 10^4$ sim units. Finally, when 50% of the total job deadline was allowed as SLA bid delay the average response time further increased to $1.956 \times 10^4$ sim units. Furthermore, in this case the end-users end up spending more budget as compared to the FCFS case (refer to Fig.4(c)).

Hence, we can see that although the proposed approach leads to better optimization of resource owners' payoff function, it has degrading effect on the end-user's QoS satisfaction function across the federation.

### 4.4.2 Resource owner perspective

In experiment 1, we quantified how varying of total SLA bid time/delay affects the individual resource owners in the federation. We analyzed, how the proposed approach affects the superscheduling parameters related to

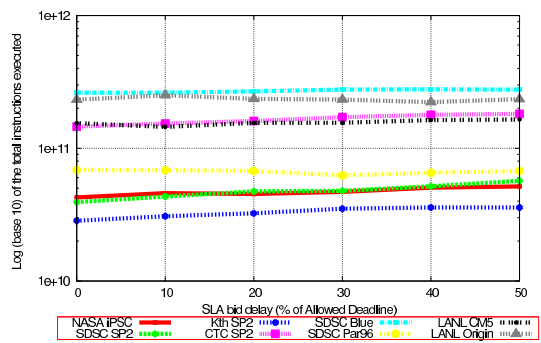

(a) total SLA bid delay vs. total MI executed

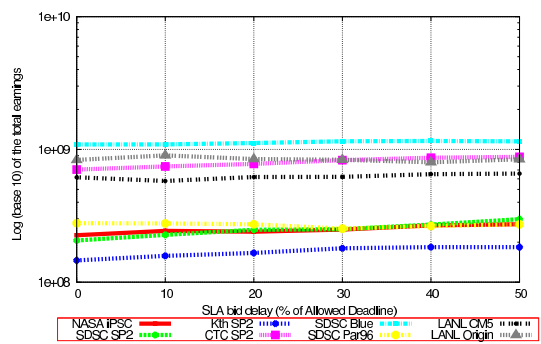

(b) total SLA bid delay vs. total earnings

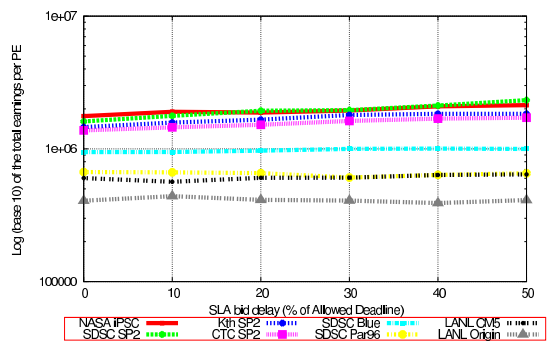

(c) total SLA bid delay vs. total earnings per processor

**Figure 5. Resource owners perspective**

the resource owner's payoff function. The most time-efficient resources in the federation i.e. NASA-iPSC, SDSC-SP2, Kth-SP2 and CTC-SP2 (refer to Table-2) experienced substantial increase in the total incentive earned with an increase in total decision making time. When no time was allowed for decision making (FCFS case), these resources earned $1.764 \times 10^6$, $1.61 \times 10^6$, $1.458 \times 10^6$, $1.377 \times 10^6$ and $9.464 \times 10^5$ grid dollars (refer to Fig. 5(c)) per processor. When the jobs in the system were allowed 30% of their total deadline as SLA bid time or admission control decision making time, these resources earned $1.946 \times 10^6$, $1.957 \times 10^6$, $1.799 \times 10^6$, $1.622 \times 10^6$ and $9.996 \times 10^5$ grid dollars per processing unit. Same trends can be observed in the plots for total earnings (refer to Fig. 5(b)) and number of machine instructions executed during the simulation period (refer to Fig. 5(a)).

Thus, we can see that when LRMSes are given decision making time, they have better control over resource allocation/admission control decision. Furthermore, we can see that Greedy backfilling approach leads to better optimization of owner's payoff function as compared to the FCFS approach.

### 4.4.3 End-users perspective

In experiment 1, we also quantified the QoS satisfaction parameters for end-user's across all the resources in the Grid-Federation. When LRMSes across the federation applied FCFS scheduling, end-users at the resource NASA-iPSC experienced $1.719 \times 10^3$ sim units as average response time (refer to Fig.6(a)). They also spent $1.143 \times 10^5$ grid dollars on the average to get their job done in the federation (refer to Fig.6(b)). However, when the user's allowed 50% of the total job deadline as SLA bid time, the average response time at NASA-iPSC increased to $3.170 \times 10^3$ sim units. In this case, end-users paid $1.14928 \times 10^5$ grid dollars. Fig.6(c) depicts the plot for number of jobs accepted for users across resources in the federation with increasing SLA bid time.

Thus, we can see that FCFS based LRMS SLA contract allocation approach is better as far as end-user's QoS satisfaction is concerned as compared to Greedy backfilling. However, such an approach is difficult to realize into today's Internet based system where resource owners have rational goals and focus on maximizing their payoff function, while delivering an acceptable level of QoS to the end-users.

### 4.4.4 System message complexity perspective

In experiment 2, we quantified the message complexity involved with our proposed superscheduling approach. We measure the number of SLA bid messages required

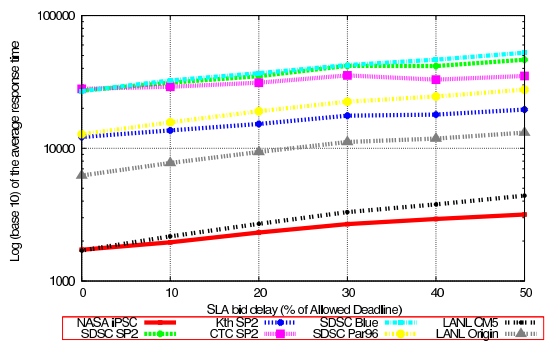

(a) total SLA bid delay vs. average response time (sim units)

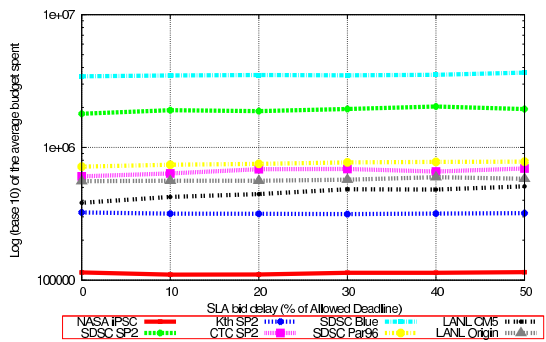

(b) total SLA bid delay vs. average budget spent (grid dollars)

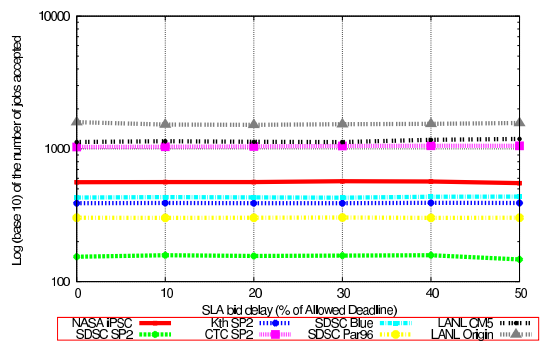

(c) total SLA bid delay vs. no. of jobs accepted

**Figure 6. End-users perspective**

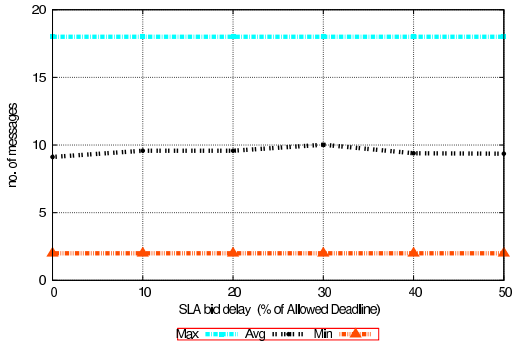

(a) total SLA bid delay vs. average no. of messages per job

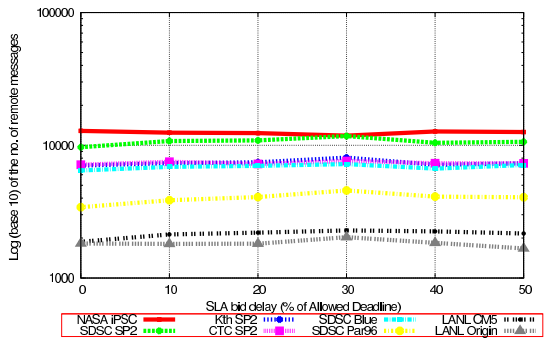

(b) total SLA bid delay vs. no. of remote messages

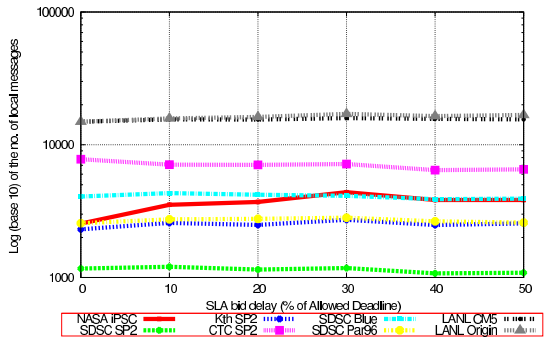

(c) total SLA bid delay vs. no. of local messages

**Figure 7. System message complexity perspective**

on average across federation to schedule a job. This metric also includes the messages for sending the executable and receiving the output. Fig.7(a), (b) and (c) depicts the plots for scheduling message complexity involved with our approach.

Our simulations show that when no SLA bid delay was allowed, the average SLA bid message per job across federation was 9.12 (refer to Fig.7(a)). As the system allowed 40% of total job deadline as SLA bid delay, the SLA message/job remained almost the same at about 9.32. Thus, we can see that our proposed superscheduling approach does not incur any additional communication overhead.

In Fig.7(b), we quantify the remote superscheduling message complexity at various resources in the Grid-Federation. We observed that the most time-efficient resource i.e. NASA-iPSC received the maximum number of remote messages followed by SDSC-SP2 and KTH-SP2. The same characteristic holds for all cases i.e. as the total SLA bid time increases from 0% to 50% of the allowed job deadline.

In Fig.7(c), we quantify the local superscheduling message complexity at various resources in the Grid-Federation. Results show that the resources LANL-Origin and LANL-CM5 were subjected to maximum local superscheduling messages. Both resources are cost-efficient and all their local users are seeking OFT. Hence these resources undertook SLA bid negotiation with time-efficient resources hence causing large number of superscheduling messages (note that number of jobs at resource LANL-Origin and LANL-CM5 were 1706 and 1287).

## 5 Related work

In this section, we briefly summarize the Grid superscheduling approaches that applies SLA-based or negotiation-based job scheduling process.

The work in [32] proposes a multi-agent infrastructure that applies SLA protocol for solving the Grid superscheduling problem. The SLA negotiation protocol is based on the Contract Net Protocol [40]. The system models three types of agents: User agent (UA), Local Scheduler agent (LSA), and Superscheduler agent (SSA). Every active site in the system instantiates these agents. The UAs are the resource consumers that submits jobs to SSA for execution on the grid platform. The UA also specifies SLA based QoS parameter such as expected response time, budget and preferred host associated with the job. The LS functionality is similar to LRMS, managing job execution within a administrative domain. LSA obtains job from the SSA that are submitted by local and remote UAs. The SSA agents are responsible for coordinating job su-perscheduling across different site in the system. The SSA agents negotiates SLA parameters with the local LSA and remote SS before scheduling the job. The model defines two kinds of SLAs: Meta-SLA and Sub-SLA. Meta-SLA refers to the initial SLA parameters submitted by the UA to its SSA. A Meta-SLA presents high-level job requirements and it can be refined during negotiation process with the SSA. While the Sub-SLA refers to the SLA parameters that are negotiated between SSA and remote site SSA. The SSA decomposes the Meta-SLAs to form Sub-SLAs. The Sub-SLA can contain much low-level resource description such as amount of physical memory required, number of processors required. In contrast we propose (i) SLA-based coordination scheme based on computational economy; and (ii) our work considers site autonomy issues, and proposes Greedy-backfilling resource allocation for a LRMS to maximize resource provider payoff function.

The work in [44] presents a grid superscheduling based on multiple job SLA negotiation scheme. The key factor motivating this work is redundantly distributing job execution requests to multiple sites in the grid instead of just sending to most lightly loaded one. The authors argue that placing job in the queue at multiple sites increases the probability that backfilling strategy will be more effective in optimizing scheduling parameters. The superscheduling parameters include resource utilization and job average turn around time. In other words, the scheduling parameters are system centric. The LRMSes at various grid sites apply FCFS policy with Easy backfilling approach for resource allocation. Further, the system proposes dual queuing system at each site. One queue for local jobs while other queue for remote jobs. During easy backfilling the local job queue is given priority over remote job queue. In contrast to this superscheduling system, our approach differs in the following: (i) the job-migration or SLA-based coordination is based on the user centric scheduling parameters; (ii) our approach gives a LRMSes more flexibility over resource allocation decision; and (iii) our cluster resource allocation mechanism i.e. Greedy backfilling algorithm focuses on maximizing resource owners payoff function.

The work in [38] models a grid superscheduler architecture. Each grid site has a grid scheduler (GS), grid middleware (GM) and a local scheduler (LRMS). Three different coordinated superscheduling scheme is presented for distributed load-balancing. These algorithms are referred to as (i) Sender-Initiated (S-I); (ii) Receiver-Initiated (RI); and (iii) Symmetrically-Initiated (Sy-I). The S-I coordination scheme is based on the information pull model, the GS sends a SLA negotiation enquiry message to all GSes in the system through its GM. Following this all contacted GSes reply with he expected average waiting time (AWT), expected run time (ERT) for the requested job and current resource utilization status (RUS). After receiving the response, the site which guarantees minimum TC (turn around cost) is preferred for job-migration. TC is computed as the sum of AWT and ERT. However, in case two GS have the same value for TC, then RUS is utilized as a tie-breaker. While the R-I approach is based on information push model. Every GS periodically checks its own RUS at time interval $\sigma$. if the RUS is below a certain predefined threshold $\delta$ then the GS volunteers itself for job-migration. It broadcasts its RUS parameter to all GSes in the system. In case, a GS needs to migrate its local job then it initiates S-I base job migration with the volunteer nodes. Finally, the Sy-I approach works in both active and passive mode. Under this approach both S-I and R-I based job migration algorithm can be initiated by the GSes in the system. Effectively, the information coordination is based on complete broadcast communication approach that may generate a large number of network messages. Such approaches has serious scalability concerns. Further, each GS in the system allocates resources to the remote and local jobs in the FCFS manner without considering any site-specific objective function. In contrast to this superscheduling system, our approach differs in the following: (i) the SLA coordination in Grid-Federation is based on one-to-one SLA negotiation mechanism hence effectively limiting the network communication overhead; and (ii) we apply Greedy backfilling approach at grid sites for maximizing resource owner payoff function.

The work in [11] presents a superscheduling system that consists of Internet-wide Condor work pools. They utilize Pastry routing substrate to organize and index the Condor work pool. Various condor pools coordinate load-balancing information by sending resource status query to all the resources in the routing table. Effectively, the query is broadcasted to all the pools that are indexed by the routing table. The resource status query (such as queue lengths and average pool utilization) includes enquiry about a pool's resource availability status and its willingness to accept remote jobs. Contacted pool manager reply with the requested info. A superscheduling manager or pool manager in the flock periodically compares the metrics such as queue lengths, average pool utilization and resource availability scenario, and based on these statistics a sorted list of pools from most suitable to least suitable is formulated. Using this list, the pool manager chooses appropriate pools for flocking. Further, a condor work pool accepts a remote job if it has free resources. The issues related to site specific resource allocation policy is not considered. In contrast: (i) we consider the site autonomy issues through Greedy backfilling LRMS scheduling approach; (ii) our SLA bidding approach gives resource owner more control before finalizing the SLA agreements; (iii) we consider one-to-one SLA coordination mechanism for superscheduling, hence largely limiting the network communication overhead; and (iv) our approach incorporates an economic mechanism for superscheduling.

Tycoon [28] is a distributed market-based resource allocation system. Job scheduling and resource allocation in Tycoon is based on decentralized isolated auction mechanism. Every resource owner in the system runs its own auction for his local resources. Furthermore, auctions are held independently, thus clearly lacking any coordination. Tycoon system relies on centralized Service Location Services (SLS) for index resource auctioneers' information. Auctioneers register their status with the SLS every 30 seconds. In case, a auctioneer fails to update its information within 120 seconds then SLS deletes its entry. Application level superschedulers contact the SLS to gather information about various auctioneers in the system. Once this information is available, the superschedulers (on behalf of users) issue bids for different resources (controlled by different auctions) constraint to resource requirement and available budget. In this setting, various superschedulers might end up bidding for small subset of resources while leaving other underutilized. In other words, superscheduling mechanism clearly lacks coordination. A resource bid is defined by the tuple $(h, r, b, t)$ where $h$ is the host to bid on, $r$ is the resource type, $b$ is the number of credits to bid, and $t$ is the time interval over which to bid. Auctioneers determine the outcome by using bid-based proportional resource sharing economy model. In contrast: (i) our superscheduling approach is based on decentralized commodity markets; and (ii) we consider a Greedy backfilling resource allocation heuristic for LRMSes.

## 6 Conclusion and future work

In this paper, we presented an SLA-based superscheduling approach based on contract net protocol. The proposed approach models set of resource providers as a contract net while job superschedulers work as managers, responsible for negotiating SLA contracts and job superscheduling in the net. Superschedulers bid for SLA contracts in the net with focus on completing the job within the user specified deadline. We analyzed how the varying degree of SLA bidding time (i.e. admission control decision making time for LRMSes) affects the resource providers' payoff function. Results show that the proposed approach gives resource owners finer control over resource allocation decisions. However, results also indicate that proposed approach has degrading effect on the user's QoS satisfaction. However, we need to do more research on abstracting the user's QoS requirement. We need to analyze how the deadline type for the user jobs can be abstracted into different types such as into urgent and relaxed deadline. In these cases, jobs with urgent requirement can be given preferences while finalizing SLA contracts hence providing improved QoS satisfaction to users.

We analyzed how the varying bid time for SLA contracts affects the system scalability and performance in terms of total message complexity. In general, the proposed superscheduling heuristic does not incur excessive messages on per job basis as compared to the FCFS case. In our future work we will study to what extent the user profile can change and how pricing polices for resources leads to varied utility of the system. We also intend to look into simultaneously bidding for SLA contracts at multiple contractor in the net, for a superscheduling iteration $l$ for a job $J_{i,j,k}$. This approach can increase the end-user's QoS satisfaction in terms of response time. As in this case the total waiting time per SLA bid is greatly reduced.

## References

[1] *http://www.cs.huji.ac.il/labs/parallel*.

[2] *http://www,hp.com/techservers/grid*.

[3] *http://www.ibm.com/grid*.

[4] *http://www.platform.com/products/wm/LSF*.

[5] *http://www.sun.com/service/utility*.

[6] D. Abramson, R. Buyya, and J. Giddy. A computational economy for grid computing and its implementation in the Nimrod-G resource broker. *Future Generation Computer Systems (FGCS) Journal, Volume 18, Issue 8, Pages: 1061-1074, Elsevier Science, The Netherlands, October*, 2002.

[7] A. O. Allen. *Probability, Statistics and Queuing Theory with computer science applications*. Academic Press, INC., 1978.

[8] A. Andrzejak and Z. Xu. Scalable, efficient range queries for grid information services. In *P2P'02: Second IEEE International Conference on Peer-to-Peer Computing*, Linkkoping, Sweden, 2002. IEEE.

[9] K. A. Berman and J. L. Paul. *Fundamentals of Sequential and Parallel Algorithms*. PWS Publishing Company, 1997.

[10] B. Bode, D. Halstead, R. Kendall, and D. Jackson. PBS: The portable batch scheduler and the maui scheduler on linux clusters. *Proceedings of the 4th Linux Showcase and Conference, Atlanta, GA, USENIX Press, Berkley, CA, October*, 2000.

[11] A. Raza Butt, R. Zhang, and Y. C. Hu. A self-organizng flock of condors. In *SC '03: Proceedings of the 2003 ACM/IEEE conference on Supercomputing*, Washington, DC, USA, 2003. IEEE Computer Society.

[12] R. Buyya, D. Abramson, J. Giddy, and H. Stockinger. Economic models for resource management and scheduling in grid computing. *Special Issue on Grid computing Environment, The Journal of concurrency and Computation:Practice and Experience (CCPE), Volume 14, Issue 13-15, Wiley Press*, 2002.

[13] R. Buyya and M. Murched. Gridsim: A toolkit for the modeling and simulation of distributed resource management and scheduling for grid computing. *Journal of Concurrency and Computation: Practice and Experience;14(13-15), Pages:1175-1220*, 2002.

[14] M. Cai, M. Frank, J. Chen, and P. Szekely. Maan: A Multi-atribute addressable network for grid information services. *Proceedings of the Fourth IEEE/ACM International workshop on Grid Computing;Page(s):184 - 191*, 2003.

[15] S. Chapin, J. Karpovich, and A. Grimshaw. The legion resource management system. *Proceedings of the 5th Workshop on Job Scheduling Strategies for Parallel Processing, San Juan, Puerto Rico, 16 April, Springer:Berlin*, 1999.

[16] J. Q. Cheng and M. P. Wellman. The WALRAS Algorithm: A convergent distributed implementation of general equilibrium outcomes. In *Computational Economics, Volume 12, Issue 1*, pages 1 – 24, Aug 1998.

[17] TH Cormen, CE Leiserson, RL Rivest, and C Stein. *Introduction to Algorithms, Second Edition*. The MIT Press, 2001.

[18] C. Courcoubetis and V. Siris. Managing and pricing service level agreements for differentiated services. In *Proc. of 6th IEEE/IFIP International Conference of Quality of Service (IWQoS'99), London, UK, May-June*, 1999.

[19] K Czajkowski, S Fitzgerald, I Foster, and C Kesselman. Grid information services for distributed resource sharing. In *HPDC '01: Proceedings of the 10th IEEE International Symposium on High Performance Distributed Computing (HPDC-10'01)*, page 181, Washington, DC, USA, 2001. IEEE Computer Society.

[20] K. Czajkowski, I. Foster, and C. Kesselman. Agreement-based resource management. In *Proceedings of the IEEE, Vol.93, Iss.3, March 2005*, Washington, DC, USA, 2005. IEEE Computer Society.

[21] D. G. Feitelson, L. Rudolph, U. Schwiegelshohn, K. C. Sevcik, and P. Wong. Theory and practice in parallel job scheduling. In *IPPS '97: Proceedings of the Job Scheduling Strategies for Parallel Processing*, pages 1–34, London, UK, 1997. Springer-Verlag.

[22] I. Foster and C. Kesselman. The grid: Blueprint for a new computing infrastructure. *Morgan Kaufmann Publishers, USA*, 1998.

[23] I. Foster, C. Kesselman, J. Nick, and S. Tuecke. The physiology of the grid: An open grid services architecture for distributed systems integration. *http://www.globus.org/research/papers.html*, 2002.

[24] J. Frey, T. Tannenbaum, M. Livny, I. Foster, and S. Tuecke. Condor-G: A computation management agent for multi-institutional grids. In *10th IEEE International Symposium on High Performance Distributed Computing (HPDC-10 '01), 2001*, pages 237 – 246, Washington, DC, USA, 2001. IEEE Computer Society.

[25] G. Gambosi, A. Postiglione, and M. Talamo. Algorithms for the relaxed online bin-packing model. *SIAM J. Comput.*, 30(5):1532–1551, 2001.

[26] W. Gentzsch. Sun grid engine: Towards creating a compute power grid. *Proceedings of the First IEEE/ACM International Symposium on Cluster Computing and the Grid*, 2002.

[27] A. Iamnitchi and I. Foster. A peer-to-peer approach to resource location in grid environments. pages 413–429, 2004.

[28] K. Lai, B. A. Huberman, and L. Fine. Tycoon: A distributed market-based resource allocation system. *Technical Report, HP Labs*, 2004.

[29] J. Litzkow, M. Livny, and M. W. Mukta. Condor- a hunter of idle workstations. *IEEE*, 1988.

[30] A. Luther, R. Buyya, R. Ranjan, and S. Venugopal. Peer-to-peer grid computing and a .net-based alchemi framework, high performance computing: Paradigm and infrastructure. 2004.

[31] D. Moore and J. Hebeler. *Peer-to-Peer:Building Secure, Scalable, and Manageable Networks*. McGraw-Hill Osborne, 2001.

[32] D. Ouelhadj, J. Garibaldi, J. MacLaren, R. Sakellariou, and K. Krishnakumar. A multi-agent infrastructure and a service level agreement negotiation protocol for robust scheduling in grid computing. In *Proceedings of the European Grid Conference*. Lecture Notes in Computer Science, Springer-Verlag, 2005.

[33] R. Ranjan, R. Buyya, and A. Harwood. A case for cooperative and incentive based coupling of distributed clusters. In *Proceedings of the 7th IEEE International Conference on Cluster Computing (CLUSTER'05), Boston, MA*.

[34] R. Ranjan, A. Harwood, and R. Buyya. Grid-federation: A resource management model for cooperative federation of distributed clusters. *Technical Report, GRIDS-TR-2004-10, Grid Computing and Distributed Systems Laboratory, University of Melbourne, Australia*, 2004.

[35] G. Sabin, R. Kettimuthu, A. Rajan, and P. Sadayappan. Scheduling of Parallel Jobs in a Heterogeneous Multi-site Environment. In *Job Scheduling Strategies for Parallel Processing, vloume 2862*, Seattle, WA, USA, june.

[36] C. Schmidt and M. Parashar. Flexible information discovery in decentralized distributed systems. *In the Twelfth International Symposium on High Performance Distributed Computing (HPDC-12), June*, 2003.

[37] J.M. Schopf. Ten actions when superscheduling. In *Global Grid Forum*, 2001.

[38] H. Shan, L. Oliker, and R. Biswas. Job superscheduler architecture and performance in computational grid environments. In *SC '03: Proceedings of the 2003 ACM/IEEE conference on Supercomputing*, page 44, Washington, DC, USA, 2003. IEEE Computer Society.

[39] S. Smale. Convergent process of price adjustment and global newton methods. In *American Economic Review, 66(2)*, pages 284–294, May 1976.

[40] R. Smith. The contract net protocol: high level communication and control in distributed problem solver. In *IEEE Transactions on Computers*, pages 1104–1113, Washington, DC, USA, 1980. IEEE Computer Society.

[41] R. G. Smith. The contract net protocol: high-level communication and control in a distributed problem solver. pages 357–366, 1988.

[42] S. Srinivasan, R. Kettimuthu, V. Subramani, and P. Sadayappan. Selective reservation strategies for backfill job scheduling. In *JSSPP '02: Revised Papers from the 8th International Workshop on Job Scheduling Strategies for Parallel Processing*, pages 55–71, London, UK, 2002. Springer-Verlag.

[43] M. Stonebraker, R. Devine, M. Kornacker, W. Litwin, A. Pfeffer, A. Sah, and C. Staelin. An economic paradigm for query processing and data migration in maiposa. *Proceedings of 3rd International Conference on Parallel and Distributed Information Systems, Austin, TX, USA, September 28-30, IEEE CS Press*, 1994.

[44] V. Subramani, R. Kettimuthu, S. Srinivasan, and P. Sadayappan. Distributed job scheduling on computational grids using multiple simultaneous requests. In *11th IEEE International Symposium on High Performance Distributed Computing (HPDC-11), 23-26 July*, Washington, DC, USA, 2002. IEEE Computer Society.

[45] C. Waldspurger, T. Hogg, B. Huberman, J. Kephart, and W. Stornetta. Spawn: A distributed computational economy. *IEEE Transactions on Software Engineering , Vol. 18, No.2, IEEE CS Press, USA, February*, 1992.

[46] J. B. Weissman and A. Grimshaw. Federated model for scheduling in wide-area systems. *Proceedings of the Fifth IEEE International Symposium on High Performance Distributed Computing (HPDC), Pages:542-550, August*, 1996.

[47] R. Wolski, J. S. Plank, J. Brevik, and T. Bryan. G-commerce: Market formulations controlling resource allocation on the computational grid. In *IPDPS '01: Proceedings of the 15th International Parallel & Distributed Processing Symposium*, page 46, Washington, DC, USA, 2001. IEEE Computer Society.

[48] C.S Yeo and R. Buyya. Service level agreement based allocation of cluster resources: Handling penalty to enhance utility. In *Proceedings of the 7th IEEE International Conference on Cluster Computing (CLUSTER'05), Boston, MA*.